\documentstyle[10pt,epsf,aaspp4]{article}
\lefthead{Formal two dimensional gravity}
\righthead{S.Engineer, K. Srinivasan and T.Padmanabhan}

\begin{document}

\parindent=0pt
\title{A formal analysis of  two dimensional gravity.}
\author{S.~Engineer$^1$, K.~Srinivasan$^1$ and T.Padmanabhan$^1$}
\affil{$^1$Inter-University Center for Astronomy and Astrophysics,\\
Post Bag 4, Ganeshkhind, Pune 411 007, INDIA}
\authoremail{srini@iucaa.ernet.in, sunu@iucaa.ernet.in and 
paddy@iucaa.ernet.in}

\begin{abstract}
\noindent Several investigations in the study of cosmological structure
formation use  numerical simulations in both two
and three dimensions.  In this paper we  address the subtle question of ambiguities  in the nature of  two  dimensional gravity in an expanding background. We take a detailed and formal approach by deriving the equations
describing gravity in $(D+1)$ dimensions using the action principle of Einstein. We then consider the Newtonian limit of these equations and finally obtain the necessary fluid equations required to describe structure formation.  These equations are solved for the density perturbation in both the linearised form  and in the spherical top hat model of nonlinear growth.  We find that, when the
special case of $D=2$ is considered, no structures can grow. We therefore conclude that, within the frame work of Einstein's theory of gravity in $(2+1)$ dimensions, formation of structures cannot take place.
Finally, we indicate the different possible ways of getting around
this difficulty so that growing structures can be obtained in two dimensional cosmological gravitational simulations  and discuss their implications. 
\end{abstract}


\section{Introduction}
\label{sec:intro}
\noindent The dominant paradigm for the generation of the observed large scale structure in the universe is based on the idea that the gravitational instability amplifies
the initially small density perturbations. The equations describing
the growth of density perturbations in the highly nonlinear stage are
analytically intractable and hence large scale numerical simulations
are resorted to for exploration of this regime. 
\par
These N--Body simulations require  large amount of computing resources
(CPU, memory and storage space) if one is to get the requisite amount of
dynamical range, {\it i.e.} good resolution in force and mass,  large range
in values of density  etc. Time and resource constraints usually
limit our ability to probe structure formation issues more deeply using computers, once the required resources are at the limits of technological feasibility. 
The key parameter which decides the feasibility level of numerical
simulations is the size of a simulation, which --- in turn --- is characterised
by:  (i)~The number of particles in the simulation volume, which is
generally specified as $N^D$, where  $D$ is the dimensionality of the
simulation (usually $2$ or $3$) and (ii)~The number
of mesh points ($M$) along any axis which determines the minimum length
scale at which the results can be treated as reliable indicators of
physical phenomena. In order to create a simulation volume that is a
fair sample of the universe one needs about  $10^7$ particles and in
order to have a high enough force resolution one needs to increase the
number of grid points adequately (for a review, see
eg. \cite{nbodyrev}). Let us suppose we have
 $160^3$ particles in three dimensions and our grid is $160$ units on a side. Then, for the same amount of computational resources  one can
simulate a two dimensional  situation with $2048^2$ particles on a $2048^2$
grid ($160^3 \approx 2048^2$). So, if we can  extract useful
(i.e. generalisable to the three dimensional case) physical insights from results in two dimensions,
then simulations of two dimensional gravity will be helpful. This hope  has led to a large number of two dimensional simulations in the field of gravitational
clustering (see eg \cite{jassunpad}, \cite{valiniaetal}, \cite{sathyaetal}, \cite{alimietal}, \cite{shandetal}).
\par
There are three ways in which two dimensional gravity can be operationally defined
and corresponding numerical simulations undertaken: (i)~Consider a
system of point particles in a three dimensional (expanding) background with the force of
interaction being given by Newton's law of gravitation ({\it i.e} $F\propto 1/r^2$). The initial positions and velocities of the
particles are such that they all lie in the same plane and all the
velocities are in the plane {\it i.e} there are no velocity components
orthogonal to the plane. This system  will evolve
with the particles being confined to the plane with clustering occurring
in the plane. Thus, we have a two dimensional clustering scenario.
(ii)~Another system we can consider consists of infinite, thin
`needles' located parallel to each other. The mass elements in the
`needles' still interact through the $1/r^2$ force, but the interaction
between `needles' (obtained by summing over the mass elements) is given by a $1/r$ force. In this case as well, the
background space expands uniformly in three dimensions. The two dimensional clustering that we study is the clustering of these `needles', examined by
taking a slice orthogonal to the `needles'. (iii)~The third possibility
involves writing down the Einstein's equations in two dimensions,
finding the homogeneous and isotropic cosmological solution, taking
the Newtonian limit (in which  the potentials due to density perturbations
and background metric can be  superposed), finding the
corresponding perfect fluid equations and solving them. In this case,
we will have a background spacetime expanding in {\it two} dimensions unlike
the other two cases.
(There is yet another, fourth possibility, which  can be defined only in an {\it ad hoc} manner. We will present this in the end.)
\par 
The first case is not of much interest for cosmological simulations
since the system is anisotropic, confined to a single plane and the
clustering takes place in a specific plane {\it only because the initial
conditions were specifically selected} to give this result. Hence we
will not discuss it and it is mentioned here only for completeness. The way  simulations in two dimensions are carried out usually is by simulating the second case and then defining
the `particles' as the intersection of the `needles' with any plane
orthogonal to them. In this case -- as in the first case -- the background
spacetime expands in three dimensions (for a flat dust dominated universe
the scale factor $a(t)$ goes as $t^{2/3}$). The clustering that we
observe and quantify in two dimensions, is basically the clustering of these
needles in three dimensions. But this is also an  anisotropic situation since
the background spacetime is expanding in three dimensions. 
As an alternative we may try to write down the equations derived from
Einstein's equations in two dimensions and examine how a system of
particles interacting in a two dimensional expanding background spacetime is to be simulated.
\par
In the rest of this paper we shall examine  the third alternative.
We will take a very general approach by developing the formal theory
of $(D+1)$ gravity and considering $D=2$  as a special case. (For some previous work on 2D gravity see eg. \cite{desjhooft} and references cited therein).
\par
The basic layout of the paper is as follows: 
In section~(\ref{sec:Dgravity}) we first define the analogue of Einstein's gravity in $(D+1)$ dimensions, discuss the Newtonian limit and its corresponding Poisson's equation and then go on to analyse the Friedmann metric in $(D+1)$ dimensions for a flat universe with dust. 
In section~(\ref{sec:structure}) we write down  the $D$ dimensional fluid equations and obtain the equation governing the density perturbations. This equation is then solved in the linear approximation and using the Spherical Top Hat model. 
Then in section~(\ref{sec:3D}) and section~(\ref{sec:2D}) we specialise to the cases $D=3$ and $D=2$ respectively. 
Finally, in section~(\ref{sec:conclusions}) we summarise and discuss the implications of the results obtained in the earlier sections.  

\section{Formal $(D + 1)$ dimensional gravity} \label{sec:Dgravity}
\noindent We start our  analysis of $(D + 1)$ dimensional ($1$ time dimension and $D$ space dimensions) gravity from the action principle which we assume has the same form as that used in $(3+1)$ dimensions.   Using this action we  construct the corresponding $(D+1)$ dimensional Einstein equations  which will be subsequently used to study structure formation and spherical collapse.  
Thus, we begin with the action principle,
\begin{equation}
{\cal S} = {\cal S}_{\rm g} + {\cal S}_{\rm m} =
-\frac{c^4}{2\kappa(D)}\int \! d^{(D+1)}x \; R \sqrt{|g|} \; + \; \int
\! d^{(D+1)}x \; {\cal L}_{\rm m} 
\label{eqn:action1}
\end{equation}
where ${\cal S}_{\rm g}$ is the action for the gravitational field, ${\cal S}_{\rm m}$ is the action for the matter fields, $g$ is the determinant of the metric tensor $g_{ik}$, $R$ is the Ricci scalar, $\kappa (D)$ is  a suitable constant which can be, in general, a function of $D$ (when $D=3$, $\kappa=8\pi G$, $G$ being the usual gravitational constant) and ${\cal L}_{\rm m}$ is the lagrangian density for the matter fields. The metric signature we adopt is $(+,-,-,-,\ldots,-)$.  We adopt the following convention regarding indices. Latin alphabets $i,j,k\ldots$ are used to represent $(D+1)$ dimensional indices which take on the values $(0,1, 2, \ldots, D)$ while greek letters are used to denote $D$ dimensional indices taking on the values $(1, 2, \ldots, D)$. Varying the total action ${\cal S}$ with respect to $g_{ik}$ we obtain Einstein's equations, 
\begin{equation}
G_{ik}\equiv R_{ik}-\frac{1}{2}\;g_{ik}\;R = {\kappa(D) \over c^4}
T_{ik} 
\label{eqn:einstein}
\end{equation}
where $T_{ik}$ is the energy momentum tensor of the matter fields and is defined by 
\begin{equation}
\frac{1}{2} \sqrt{|g|} T_{ik} = \frac{\partial (\sqrt{|g|}{\cal L}_{\rm m})} {\partial g^{ik}} \; - \; 
\frac{\partial}{\partial x^l} \left( \frac{\partial (\sqrt{|g|}{\cal
L}_{\rm m})}{\partial \left( \partial g^{ik}/\partial x^l \right) }
\right)
\label{eqn:emtensor}
\end{equation}
$G_{ik}$ is the Einstein tensor and $R_{ik}$ is the usual Ricci
tensor. Note that the $(1/2)$ that appears in Einstein's equations
arises due to the square root in the term $\sqrt{|g|}$ and has nothing to do with the dimension of the spacetime.  
\par
We will use the above equations in the subsections to follow.  
In subsection~(\ref{subsec:poisson}), we will study the Newtonian limit of the metric tensor and then construct the corresponding Poisson equation that relates the Newtonian gravitational field $\phi$ to the matter density $\rho$.  
Then, in subsection~(\ref{subsec:friedmann}), we analyse the Friedmann metric in $(D+1)$ dimensions and the corresponding Newtonian limit of this metric is derived.  
\subsection{Poisson equation in $D$ dimensions}
\label{subsec:poisson} 
\noindent In this section, we derive the Poisson equation relating the
gravitational potential $\phi$ to the matter density $\rho$. We keep
all factors of $c$ since the Newtonian limit involves the limit $c\to
\infty$. The analysis here follows closely the treatment in \cite{landau2}.
  Consider the metric.
\begin{equation}
ds^2 = \left( 1 + {2\phi \over c^2} \right)c^2dt^2 - dl^2
\label{eqn:Nmetric}
\end{equation}
where $\phi$ is a function of space and time with dimensions of velocity square. The
term  $dl^2$ is the $D$ dimensional spatial line element given by the formula
\begin{equation}
dl^2 = \sum_{\alpha=1}^D (dx^{\alpha})^2 
\label{eqn:Dline}
\end{equation}
We will now show that the metric written above is the Newtonian limit of Einstein's gravitational equations. We do this by showing that, in the Newtonian limit, the equation of motion of a particle follows Newton's force law with the force $-m\nabla \phi$...  
In relativistic mechanics, the motion of a particle of mass $m$ is determined by the action function $S$
\begin{equation}
S = -mc \int \! ds = -mc \int \! c\,dt \sqrt{\left( 1 + {2\phi
\over c^2}  -{v^2 \over c^2}\right)}
\label{eqn:Nmetric1}
\end{equation}
where $v^2$ is the square of the magnitude of the particle's velocity in $D$ dimensions. In arriving at the second equality we have used the form of the metric in equation~(\ref{eqn:Nmetric}). In the limit $c \to \infty$, the action $S$ can be approximated as
\begin{equation}
S \approx -mc^2 \int \! dt \left( 1 + {2\phi -v^2 \over 2c^2} \right)
= \int \! dt \left( -mc^2 + {1 \over 2}mv^2 - m\phi\right)
\label{eqn:Nmetric2}
\end{equation}
The equation of motion for the particle can be immediately written down and we obtain
\begin{equation}
m{d{\bf v} \over dt} = - m\nabla \phi 
\label{eqn:Nmetric3}
\end{equation}
where ${\bf v}$ is the velocity vector in $D$ space dimensions.  Thus, Newton's force law is recovered in the non-relativistic limit and from this we conclude that the metric given in equation~(\ref{eqn:Nmetric}) is the Newtonian limit of Einstein's gravitational equations with $\phi$ acting as the Newtonian gravitational potential. 

The relation between $\phi$ and the mass density $\rho$ is found by taking the $c\to\infty$ limit of
Einstein's equations. This procedure, in $(3+1)$ gravity, determines the constant $\kappa(D)$ since the Poisson equation is explicitly known. In other dimensions however, a definite criterion, like Gauss's law for example, must be imposed in order to determine $\kappa(D)$. We now consider the limit $c \to \infty$ of Einstein's equations in the following manner.
First, we use the line element given in equation~(\ref{eqn:Nmetric}) to calculate the Ricci tensor component $R_{00}$
\begin{equation}
R_{00} = \frac{1}{c^2}\frac{1}{c^2 + 2 \phi} (\partial_\mu \phi)(\partial^\mu \phi) - \frac{1}{c^2} \partial_\mu\partial^\mu \phi 
\label{eqn:poisson1}
\end{equation}
where the summation convention has been invoked in the above equation and the sum over $\mu$ is only over the {\it spatial} dimensions. 
Then, using equation~(\ref{eqn:einstein}), we obtain,
\begin{equation}
R = -{2\kappa(D) \over c^4(D-1)} T
\end{equation}
where we have used the fact that $g_{ik}g^{ik} = D + 1$ and assumed $D \neq 1$. Thus, Einstein's equations can be written in the equivalent form,  
\begin{equation}
R_{ik} = {\kappa(D) \over c^4} \left( T_{ik} - {1\over D-1}g_{ik}T \right). \label{eqn:poisson2}
\end{equation}
where $T$ is the trace of $T_{ik}$.  
The energy momentum tensor of point particles is $T_{ik} = \rho c^2 u_iu_k$ where $\rho$ is the mass density and $u_i$ is the four velocity.  Since, in the non-relativistic limit, the macroscopic motion is slow, the space components of $u_i$ can be neglected and only the time component should be retained.  Therefore, $u_0 = \sqrt{g_{00}}$ and $u_\mu \approx 0$ for all $\mu$. Consequently, only $T_{00} = g_{00}\rho c^2$ is non-zero.  Substituting 
for $T_{ik}$ into equation~(\ref{eqn:poisson2}) and using the expression in equation~(\ref{eqn:poisson1}) for $R_{00}$, we get,  
\begin{equation}
{1 \over c^2} \partial_\mu\partial^\mu \phi = -\left({D-2 \over
D-1}\right) \left(1 + {2\phi \over c^2}\right) {\kappa(D) \over c^4}
\rho c^2 \approx  -\left({D-2 \over D-1}\right){\kappa(D)\over c^2}
\rho 
\label{eqn:poisson3}
\end{equation}
That is,
\begin{equation}
\nabla^2 \phi = \left({D-2 \over D-1}\right)\kappa(D) \rho 
\label{eqn:poisson4}
\end{equation}
where $\nabla^2$ is the usual Laplacian operator in $D$ dimensions. This equation is the Poisson equation in $D$ dimensions. Note that when substituting for the value of $R_{00}$ from equation~(\ref{eqn:poisson1}), we neglected the first term in comparison with the second since the former is of order $c^{-4}$ while the latter is only of order $c^{-2}$.  

\subsection{Friedmann Universe in $(1+D)$ dimensions} 
\label{subsec:friedmann}
\noindent Let us next consider the maximally symmetric Robertson-Walker metric in $(D+1)$ dimensions, specialising to flat space with $k=0$ (we set $c=1$ in this and in subsequent sections),
\begin{equation}
ds^2 = dt^2 - a^2(t)dl^2 
\label{eqn:robertson}
\end{equation}
where $a(t)$ is the scale factor and $dl^2$ is the $D$ dimensional line element given in equation~(\ref{eqn:Dline}). Calculating the components of the Einstein tensor, we obtain,
\begin{eqnarray}
G_{00} &=& \frac{D(D-1)\dot{a}^2}{2a^2} \nonumber \\
G_{11} = G_{22} =\ldots = G_{DD} &= & (1-D)a \ddot{a} + \left(1 -
\frac{D}{2}\right) (D-1) \dot{a}^2 
\label{eqn:fried1}
\end{eqnarray}
where $\dot{a}$ stands for $da(t)/dt$ and similarly $\ddot{a}$ is the second derivative of $a(t)$ with respect to time. All the other components are zero.  For consistency, the energy
momentum tensor must have the form $T^i_{\; k} = {\rm diag}(\rho,
-p,-p,-p,\ldots)$ where $\rho$ is the matter density and $p$ is the
pressure. 
\par
Substituting in Einstein's equations, we obtain,
\begin{eqnarray}
\frac{D(D-1)\dot{a}^2}{2a^2}& =& \kappa(D) \rho  \label{eqn:friedmann1} \\
\frac{\ddot{a}}{a} +\frac{D-2}{2} \frac{\dot{a}^2}{a^2}&=& -\frac{\kappa(D) p}{D-1} \label{eqn:friedmann}
\end{eqnarray}
The above two equations, together with the equation of state in the form $p =p(\rho)$ completely specify the system.  Solving these three equations, we can determine $a(t)$, $\rho(t)$ and subsequently $p(t)$.
Combining equations~(\ref{eqn:friedmann1},\ref{eqn:friedmann}), we get the single equation,
\begin{equation}
\frac{\ddot{a}}{a} = -\frac{\kappa(D)}{D(D-1)}\left[ (D-2)\rho + Dp\right] .\label{eqn:friedmann2}
\end{equation}
\noindent We now specialise to the case of pressureless dust with the equation of state $p =0$.  Using the principle of conservation of energy and momentum expressed by the relation
\begin{equation}
T^k_{\;\; i;k}=0 \label{eqn:fried2}
\end{equation}
we derive the following relation,
\begin{eqnarray}
T^k_{\;\; i;k} &=& {1 \over \sqrt{|g|}} {\partial \over \partial x^k}\left( \sqrt{|g|} T^k_{\;\; i} \right) - {1 \over 2}{\partial g_{kl} \over \partial x^i} T^{kl} \nonumber \\
&=& {1 \over a^D} {\partial \over \partial x^k}\left(a^D T^k_{\;\;
i}\right) - {1 \over 2}{\partial g_{kl} \over \partial x^i} T^{kl} = 0
 \label{eqn:fried3}
\end{eqnarray}
Noting that the only non-zero component of $T^i_{\; k}$ is $T^0_{\; 0} = \rho$ we finally get
\begin{equation}
\rho a^D = {\rm constant} = C_1 \label{eqn:densityrel}
\end{equation}
Substituting the above relation into equation~(\ref{eqn:friedmann1}), and solving for $a(t)$ and subsequently for $\rho(t)$, we obtain the solutions,
\begin{equation}  
a(t) = \left({ D\kappa(D) C_1 \over 2(D-1) }\right)^{1/D} \, t^{2/D} \quad ; \quad \rho(t) = \left({2 (D-1) \over D \kappa(D)}\right) t^{-2} 
\label{eqn:friedsoln}
\end{equation}
Let us next consider the Newtonian limit of the Friedmann metric. This limit is important because the length scales of interest in structure formation are small compared to the Hubble radius and the velocities in the system are also much smaller than {\it c}. This permits us to study the formation of large scale structures in the universe in a Newtonian framework where the effective potential due to the expanding background universe, $\Phi_{\rm FRW}$, and the potential due to the density perturbations, $\varphi$, can be simply superposed.  In order to obtain $\Phi_{\rm FRW}$, we first recast the Friedmann metric in equation~(\ref{eqn:robertson}) into the more convenient form
\begin{equation}
ds^2 = dt^2 - a^2(t)\left( dX^2 + X^2 d\Omega^2 \right)
\end{equation}
where $X$ is the radial distance in $D$ dimensions and $\Omega$ is the
corresponding solid angle.  We then apply the transformations (see \cite{probbook}, pg 80,346)
\begin{equation}
r = Xa(t) \quad , \quad  T = t - t_0 + {1\over 2} a\dot{a} X^2 + {\cal O}(X^4)
\end{equation}
where only terms up to quadratic in $X$ are retained.  Direct calculations, correct upto this order, transforms the Friedmann line element to the form,
\begin{equation}
ds^2 \approx \left(1 - {\ddot{a} \over a}r^2\right) dT^2 - dr^2 - r^2 d\Omega^2
\end{equation}
which upon comparison with the metric in equation~(\ref{eqn:Nmetric}) gives the equivalent Newtonian potential $\Phi_{\rm FRW}$ in $D$ dimensions as 
\begin{equation}
\Phi_{\rm FRW}=-\frac{1}{2} \frac{\ddot{a}}{a}\;r^2   
\label{eqn:bgpotential}
\end{equation}
\noindent We will now use the results developed in the last two subsections to study structure formation and spherical collapse using the STH model.

\section{Structure formation in $D$ dimensions} \label{sec:structure}
\noindent Having determined the form of the Poisson equation in the Newtonian limit and analysed the Friedmann equations in $(D+1)$ dimensions, we proceed to derive the equation for the growth of inhomogeneities in the expanding universe. After this we consider a specific model, the STH model, to study spherical collapse of matter.

\subsection{Equation for density perturbations in $D$ dimensions}  
\noindent Let us assume that matter in the universe is a perfect, pressureless fluid with density  $\rho_{\rm m}$ and  flow velocity ${\bf U}$.  We can formally write down the $D$ dimensional fluid equations describing a perfect fluid in an 
external potential field $\Phi_{\rm tot}$  in a proper coordinate system labelled by the $D$ dimensional vector ${\bf r}$. Therefore, we have,  
\begin{eqnarray}
\left(\frac{\partial\;\rho_{\rm m}}{\partial t}\right)_{\bf r}+\, \nabla_{\bf r}\cdot (\rho_{\rm m}\;{\bf U})&=& 0  \label{eqn:continuity} \\
\left(\frac{\partial \bf U}{\partial t}\right)_{\bf r}+ \, \left({\bf U} \cdot \nabla_{\bf r}\right) {\bf U} & = &-\nabla_{\bf r} \Phi_{\rm tot} \label{eqn:euler}
\end{eqnarray}
where equation~(\ref{eqn:continuity}) is the usual continuity equation while equation~(\ref{eqn:euler}) is the Euler equation for the fluid. The potential in equation~(\ref{eqn:euler}), $\Phi_{\rm tot}$, is the total external Newtonian potential 
\begin{equation}
\Phi_{\rm tot}= \Phi_{\rm FRW} + \varphi
\end{equation}
where $\Phi_{\rm FRW}$ is the background potential associated with the smooth background matter density $\rho_{\rm bm}$ and is given in  equation~(\ref{eqn:bgpotential}) 
while $\varphi$ is the potential caused by density perturbations $(\rho_{\rm m} - \rho_{\rm bm})$.  The potential $\varphi$ satisfies the Poisson equation given in  equation~(\ref{eqn:poisson4}).  Thus, 
\begin{equation}
\nabla_{\bf r}^2 \varphi = \left(\frac{D-2}{D-1}\right) \kappa(D)(\rho_{\rm m} - \rho_{\rm bm}) = \left(\frac{D-2}{D-1}\right)\kappa(D)\rho_{\rm bm}\delta \label{eqn:Poisson}
\end{equation}
where $\delta$ is the density contrast defined by 
\begin{equation}
\delta = \frac{\rho_{\rm m} - \rho_{\rm bm}}{\rho_{\rm bm}} \label{eqn:dcontrast}
\end{equation} 
We now transform to comoving coordinates defined by ${\bf x}={\bf r}/a(t)$ and define the peculiar velocity ${\bf v}$ by the relation
\begin{equation}
{\bf U} = H(t){\bf r} + {\bf v} = \dot{a}{\bf x} + {\bf v}
\end{equation}
where ${\bf v}= a\dot{{\bf x}}$ and $H(t)=(\dot{a}/{a})$.
Then, equation~(\ref{eqn:continuity}) and equation~(\ref{eqn:euler}) become  
\begin{eqnarray}
&&\left(\frac{\partial \rho_{\rm m}}{\partial t}\right)_{\bf x} + \, D H \rho_{\rm m}+\frac{1}{a}  \nabla_{\bf x}\cdot(\rho_{\rm m} {\bf v})= 0 \label{eqn:cocontinuity}\\
&&\left(\frac{\partial {\bf v}}{\partial t}\right)_{\bf x} + \, H {\bf v}+\frac{1}{a} \left( {\bf v} \cdot\nabla_{\bf x} \right) {\bf v}=-\frac{1}{a} \nabla_{\bf x} \varphi \label{eqn:coeuler}
\end{eqnarray}
where we have used equation~(\ref{eqn:bgpotential}) to substitute for $\Phi_{\rm FRW}$.
Similarly, in co-moving co-ordinates, equation~(\ref{eqn:Poisson}) reduces to 
\begin{equation}
\nabla_{\bf x}^2 \varphi =  \left(\frac{D-2}{D-1}\right)\kappa(D) a^2\rho_{\rm bm}\delta \label{eqn:coPoisson}
\end{equation}

\noindent Using $\rho_{\rm m}=\rho_{\rm bm} (1+\delta)$, transforming the time variable from $t$ to $a(t)$ and defining a new velocity variable ${\bf u}$ by 
\begin{equation}
{\bf u}= \frac{d{\bf x}}{da} = \frac{{\bf v}}{a\dot{a}}
\end{equation}
we can obtain equations for $\delta (a)$ and ${\bf u}(a)$.  Therefore, using equation~(\ref{eqn:densityrel}) and performing the transformations, equation~(\ref{eqn:cocontinuity}) and equation~(\ref{eqn:coeuler}) further reduce to, 
\begin{eqnarray}
&&\frac{\partial \delta}{\partial a}+\nabla_{\bf x}\cdot [{\bf u}(1+\delta)]=0 
\label{eqn:cocontinuity1} \\
&& \dot{a}^2 \frac{\partial {\bf u}}{\partial a} + \left(\ddot{a} + 2\frac{\dot{a}^2}{a} \right){\bf u} + \dot{a}^2 ({\bf u}\cdot \nabla_{\bf x}) {\bf u} = -\frac{1}{a^2} \nabla_{\bf x}\varphi \label{eqn:coeuler1}
\end{eqnarray}
Now, we use the Friedmann equations in equations~(\ref{eqn:friedmann1},\ref{eqn:friedmann}) with $\rho$ replaced by $\rho_{\rm bm}$ and with $p=0$ to substitute for $\ddot{a}$ in the above equation. Further, we define a new potential $\Psi$ by the relation
\begin{equation}
\Psi = \left(\frac{D(D-1)}{6-D}\right)  \frac{1}{\kappa(D) \rho_{\rm bm} a^3}\, \varphi
\end{equation} 
so that, upon using equation~(\ref{eqn:coPoisson}), one obtains, 
\begin{equation}
\nabla_{\bf x}^2 \Psi =  \left(\frac{D(D-2)}{6-D}\right) \frac{\delta}{a} \label{eqn:coPoisson1}
\end{equation}
where all reference to $\kappa(D)$ has disappeared.  Hence the final system of equations we need to tackle are, 
\begin{eqnarray}
&&\frac{\partial \delta}{\partial a}+\nabla_{\bf x}\cdot [{\bf u}(1+\delta)]=0 
\label{eqn:cocontinuity2} \\
&&\frac{\partial {\bf u}}{\partial a} + ({\bf u}\cdot \nabla_{\bf x}) {\bf u} = -\frac{6-D}{2a} A \left[ \nabla_{\bf x}\Psi + {\bf u}\right]
\label{eqn:coeuler2}
\end{eqnarray}
where $A$ is given by the relation
\begin{equation}
A = \left(\frac{2\kappa(D) }{D(D-1)}\right) \frac{a^2}{\dot{a}^2}\, \rho_{\rm bm} = \frac{\rho_{\rm bm}(t)}{\rho_c(t)} \qquad ; \qquad \rho_c \equiv \left({D(D-1) \over 2\kappa(D)}\right) {\dot{a}^2 \over a^2}
\end{equation}
For the $k=0$ universe, we will set $A=1$.  
\par
To proceed further and determine the equation satisfied by $\delta$, we decompose the term $\partial_{\alpha} u_{\beta}$ (where $u_{\beta}$ is the $\beta$th covariant component of the vector ${\bf u}$ and $\partial_\alpha$ is short for $\partial/\partial x^{\alpha}$) as
\begin{equation}
\partial_{\alpha} u_{\beta} = \sigma_{\alpha\beta} + \Omega_{\alpha\beta} + \frac{1}{D}\delta_{\alpha\beta}\theta \qquad \alpha,\beta = (1,2, \ldots, D)
\end{equation}
where $\sigma_{\alpha\beta}$ is the traceless, symmetric shear tensor,
$\Omega_{\alpha\beta}$ is the antisymmetric rotation tensor, $\theta$ is the
(trace) expansion and $\delta_{\alpha\beta}$ is the Kronecker delta symbol.
Then, equation~(\ref{eqn:cocontinuity2}) and equation~(\ref{eqn:coeuler2}) are combined by taking the divergence of equation~(\ref{eqn:coeuler2}) and using the above decomposition of $\partial_\alpha u_\beta$ to obtain a single equation for $\delta$.  Straight forward algebra gives,  
\begin{equation}
\frac{d^2 \delta}{d a^2}+ \left(\frac{6-D}{2 a}\right) \frac{d \delta}{d a}-
\left(\frac{D (D-2)}{2 a^2}\right) \delta(1+\delta)=  \left(\frac{D+1}{D}\right)\frac{1}{(1+\delta)} \left( \frac{d \delta}{d a}\right)^2 
 + \, (1+\delta) (\sigma^2-2\Omega^2) \label{eqn:ddelta}
\end{equation}
where $\sigma^2 = \sigma_{\alpha\beta}\sigma^{\alpha\beta}$ and $\Omega^2 = (1/2)\Omega_{\alpha\beta} \Omega^{\alpha\beta}$. This equation is the full non-linear equation for $\delta$. 
Apart from the obvious nonlinear terms
containing $\delta^2$ and $(d\delta/da)^2$, the term
$(1+\delta)(\sigma^2-2\Omega^2)$,  which is the contribution from the
shear and rotation, is also non-linear. The non-linear terms in $\delta$ in the above equation render the
equation unsolvable in general.  Ignoring these non-linear
terms to a first approximation, we can get a linear equation for $\delta (a)$,   \begin{equation}
\frac{d^2 \delta}{d a^2} + \left(\frac{6-D}{2 a}\right) \frac{d \delta}{d a} - \left(\frac{D (D-2)}{2 a^2}\right)\delta = 0.
\end{equation}
Assuming a power law solution for delta in the form $\delta \propto a^p$, we get, 
\begin{equation}
p = \frac{D-4}{4} \, \pm \, \frac{1}{4} \sqrt{ 9D^2 - 24D + 16}
\end{equation}
as the required values for $p$. Notice that $\delta$ has a growing
mode as well as a decaying mode in general.  The above solutions hold
for all values of $D>1$ in the linear regime.  
\par
Though the full non-linear equation is not solvable, by neglecting the contribution from the shear and rotation terms and  by using a suitable ansatz for $\delta$, the resulting non-linear equation {\it can} be solved.  We proceed to do this within the framework of the STH model in the next section.    

\subsection{The Spherical Top Hat (STH) model}
\noindent In the STH (spherical collapse) model, we assume spherical symmetry by neglecting the shear and rotation terms in the equation for $\delta$. With this assumption the $\delta$ equation can be exactly solved.
\par   
Transforming equation~(\ref{eqn:ddelta}) by changing the independent variable back to $t$, dropping the rotation and shear terms and using the Friedmann equations given in equations~(\ref{eqn:friedmann1},\ref{eqn:friedmann}), we get,
\begin{equation}
\frac{d^2 \delta}{d t^2}+ 2 \frac{\dot{a}}{a} \frac{d \delta}{d t} -\left(\frac{D+1}{D}\right)\frac{1}{(1+\delta)}\left(\frac{d \delta}{d t}\right)^2=\left(\frac{D-2}{D-1}\right) \kappa(D)\rho_{\rm bm} \delta(1+\delta) \label{eqn:deltatime}
\end{equation}  
We now define a function $R(t)$ by the relation
\begin{equation}
1+\delta=\frac{\rho}{\rho_{\rm bm}}=\frac{M}{C_D R^D (t) \rho_{\rm bm}} \label{eqn:dansatz}
\end{equation}
where  $C_D=2 \pi^{D/2}/(D \Gamma[D/2])$ is the volume of a unit sphere in $D$ dimensions introduced for later convenience and $M$ is a constant.  
The expression for $\delta$ above can be rewritten using the relation $\rho_{\rm bm}a^D = \rho_0 a_0^D$ from equation~(\ref{eqn:densityrel}):
\begin{equation}
1+\delta=\frac{M}{C_D \rho_0 a_0^D} \left[\frac{a}{R}\right]^D\\
=\lambda \frac{a^D}{R^D} \label{eqn:dansatz1}
\end{equation}
where $\rho_0$ and $a_0$ are the matter density and scale factor at
some (arbitrarily chosen) ``present" epoch $t_0$.
Substituting equation~(\ref{eqn:dansatz1}) in equation~(\ref{eqn:deltatime}), we get an equation for the growth of $R(t)$ as,
\begin{equation}
\frac{d^2R}{dt^2} = -\frac{D-2}{D(D-1)} \frac{\kappa(D)}{C_D}\frac{M}{R^{D-1}}
\end{equation}
[As an aside we may note that if the universe contains matter or fields
with equations of state other than $p=0$, the equation for $R(t)$ becomes
\begin{equation}
\frac{d^2R}{dt^2} = -\frac{D-2}{D(D-1)} \frac{\kappa(D)}{C_D}\frac{M}{R^{D-1}} \; - \; \frac{\kappa(D)}{D(D-1)}\left( (D-2)\rho + Dp\right)_{\rm rest} R
\end{equation}
where the term $((D-2) \rho+Dp)_{rest}$ comes from the smoothly distributed
component with $p\neq 0$.]
\par
>From the form of the equation of motion of $R(t)$ we can give the following interpretation. Since the entire system considered above is spherically symmetric, we interpret $R$ as the radius of a $D$ dimensional spherical region containing a mass $M$. The equation of motion of $R$ determines the motion of the surface of this region. In general, a spherical overdense region will be expected to initially expand because of the expansion of the background universe till the excess gravitational force due to the overdensity of enclosed matter stops the expansion and causes the region to collapse back on itself. We will discuss the cases $D=3$ and $D=2$ in the subsequent sections and determine the differences in the behaviour of the growth of inhomogeneities.   

\section{Summary of standard results in 3-dimensions} \label{sec:3D}
\noindent When $D=3$, all the standard equations are recovered.  First the Poisson equation satisfied by the Newtonian gravitational potential given by equation(\ref{eqn:poisson4}) reduces to the standard form,
\begin{equation}
\nabla^2 \phi = 4\pi G \rho \label{eqn:3Dpoisson}
\end{equation}
where we have defined $G$ by relating it to $\kappa(3)$ by  $\kappa(3)
= 8\pi G$. Similarly, equations (\ref{eqn:cocontinuity2}) and
equation (\ref{eqn:coeuler2}) reduce to (with $A=1$), 
\begin{eqnarray}
&&\frac{\partial \delta}{\partial a}+\nabla_{\bf x}\cdot [{\bf u}(1+\delta)]=0 
\label{eqn:3Dcocontinuity2} \\
&&\frac{\partial {\bf u}}{\partial a} + ({\bf u}\cdot \nabla_{\bf x}){\bf u} = -\frac{3}{2a}  \left[ \nabla_{\bf x}\Psi + {\bf u}\right]
\label{eqn:3Dcoeuler2}
\end{eqnarray}
while the equation for $\Psi$ becomes,
\begin{equation}
\nabla_{\bf x}^2 \Psi = \frac{\delta}{a}. \label{eqn:3DcoPoisson1}
\end{equation}
In a similar manner, the $\delta$ equation reduces to,
\begin{equation}
\frac{d^2 \delta}{d a^2}+\frac{3}{2 a} \frac{d \delta}{d a}-
\frac{3}{2 a^2}\delta(1+\delta)=\frac{4}{3 (1+\delta)} \left( \frac{d \delta}{d a}\right)^2+(1+\delta) (\sigma^2-2\Omega^2) \label{eqn:3Dddelta}
\end{equation}
and the solutions to the linear perturbation equation which is
obtained by dropping the nonlinear terms and the
$(\sigma^2-2\Omega^2)$ term, are
\begin{equation}
\delta \propto a^p,  \quad p = 1, -\frac{3}{2}
\end{equation}
which are well known.
The STH model for $D=3$ also reduces to the standard form
\begin{equation}
\frac{d^2R}{dt^2} = -\frac{G M}{R^2} \; - \; \frac{4\pi G}{3}\left(\rho + 3p\right)_{\rm rest} R
\end{equation}
which is again well known (see \cite{probbook}).
Therefore, it is seen that the full $D$ dimensional equations reduce to the correct equations in three dimensions.  Now we will go on to discuss the important case of $D=2$.

\section{Two dimensional Gravity} \label{sec:2D}
\noindent If we naively consider the limit $D \to 2$ in the $D$ dimensional equations,
assuming that $\kappa(D)$ is finite in this limit, we obtain the following
results.  First, the Poisson equation (\ref{eqn:poisson4}) reduces to
\begin{equation}
\nabla^2 \phi = 0
\label{eqn:2Dpoisson1}
\end{equation}
The above result shows that in two dimensions, the
gravitational potential does {\it not} couple to the matter density $\rho$.  In structure formation, this means that inhomogeneities cannot grow since the perturbed potential $\varphi$ is not related to  $\delta$ at all. 
The second interesting  result is that the background Newtonian potential $\Phi_{\rm FRW}$ vanishes. This occurs because, referring back to equation~(\ref{eqn:friedmann2}), $\ddot{a} = 0$ for pressureless dust and hence the background potential is zero. 
Further, the $\delta$ equation reduces to 
 \begin{equation}
\frac{d^2 \delta}{d a^2}+\frac{2}{a} \frac{d \delta}{d a}=\frac{3}{2 (1+\delta)} \left( \frac{d \delta}{d a}\right)^2+(1+\delta) (\sigma^2-2\Omega^2). \label{eqn:2Dddeltalin}
\end{equation} 
Linearising the equation as before by dropping the
$(\sigma^2-2\Omega^2)$ and $(d\delta/da)^2$ terms we obtain
\begin{equation}
\frac{d^2 \delta}{d a^2}+\frac{2}{a} \frac{d \delta}{d a}=0. \label{eqn:2Dddelta}
\end{equation} 
 The solutions to the linearised equation are 
\begin{equation}
\delta \propto a^p, \quad p =0, -1.
\end{equation}
Thus only a constant or the decaying mode is present.  This is
consistent  with the result that the perturbed gravitational potential does not couple to $\delta$. If one considers the STH model, it is easy to see that the growth equation for $R(t)$ reduces to 
\begin{equation}
\frac{d^2R}{dt^2} = -\kappa(2) p_{\rm rest} R
\end{equation}
For, $p_{\rm rest} = 0$, the solution to the above equation is just
$R(t) = B_1 t + B_2$ where $B_1, B_2$ are constants. This is to be
expected since there is no gravitational force which can lead to clustering
and as a consequence the radius simply grows with time just like the
background universe.
Thus, if $\kappa(D)$ is {\it finite} in the limit $D\to 2$,
it is not possible to have gravitational clustering
that can grow with time. 
\par
We can, however, try some alternative approaches to examine whether it is possible to
have a consistent physical picture of growing structures for two
dimensional gravity. 
\par
One possibility is that instead of assuming $\kappa(D)$ to be finite, let us assume that the expression $(\kappa(D) (D-2))$ remains finite when $D \to 2$. This finite value can be fixed, for example, by invoking Gauss's theorem in $D$ dimensions. This gives
\begin{equation}
\kappa (D)=\left(\frac{D-1}{D-2}\right)\frac{2 \pi^{D/2} G}{\Gamma [D/2]} \label{eqn:poisson5}
\end{equation}
Thus, $\kappa(D) \to \infty$ when $D\to 2$, but the Poisson equation acquires the form
\begin{equation}
\nabla^2 \phi = 2 \pi G \rho \label{eqn:2Dpoisson2}
\end{equation}
This is, of course, the same form which is obtained by applying Gauss's law in two dimensions. Hence, as in three dimensions, the gravitational potential is determined by the matter density and thus inhomogeneities can in
principle grow.
There are, however, difficulties with this approach. To begin
with, the constant factor $c^4/(2 \kappa (D))$ in the action
$\cal S$ in equation(\ref{eqn:action1}) vanishes for $D=2$.
But this is not too serious a problem.  The gravitational part of the action certainly vanishes but because only the variations about the action are of significance this difficulty can be ignored. But a more serious problem arises when the solutions to the Friedmann equation are considered.  
The solutions for $a(t)$ and $\rho(t)$ in $D$ dimensions are given in
equation~(\ref{eqn:friedsoln}). Using equation~(\ref{eqn:poisson5}), these reduce to, 
\begin{equation}  
a(t) = \left({ D\pi^{D/2} G C_1 \over (D-2) \Gamma[D/2] }\right)^{1/D}  t^{2/D}
\quad ; \quad \rho(t) = C_1 a^{-D} = { (D-2) \Gamma[D/2] \over D
\pi^{D/2} G} \, t^{-2} 
\end{equation}
When $D\to 2$ then,  $a \to \infty$ and $\rho \to 0$ irrespective of the dependence on $t$.  This implies that
one cannot solve the equations describing the growth of structure in a
consistent and non--singular way. 
Hence, we conclude that it is not possible to have a theoretical
formulation of two dimensional gravity as the Newtonian limit to
Einstein's equations in two dimensions. 
\par
An alternative that remains is to use the Newtonian fluid equations
in $D$ dimensions directly and rewrite them for an expanding
background with an {\it arbitrary}
scale factor $a(t)$.  Note that $a(t)$ is not obtained from the Friedmann equations and is completely arbitrary.  We can superpose the potentials for the
background universe and the perturbations in this case as before. The further assumptions we need to make are 
(i)~the potential of the background
universe $\Phi_{\rm bg}$ is of the form 
\begin{equation}
\Phi_{\rm bg} = -\frac{1}{2} \frac{\ddot{a}}{a}\, r^2
\end{equation} 
and (ii)~the Poisson equation is given by 
\begin{equation}
\nabla^2 \phi=\kappa(D) \rho_{\rm bm} \delta
\end{equation}
where $\kappa(D)=2 \pi^{D/2} G/\Gamma [D/2]$. This form of $\kappa(D)$ is
obtained from the use of Gauss's law in $D$ dimensions.  We also need to
specify how the background density $\rho_{\rm bm}$ depends on  time. In analogy with the usual Friedmann equations, we will assume $\rho_{\rm bm} a^D=C_1$ where $C_1$ is a constant.
This gives an equation for $\delta$ with an arbitrary scale factor
$a(t)$ as
\begin{equation}
\frac{d^2 \delta}{d a^2}+\left(\frac{\ddot{a} a+2 \dot{a}^2}{a \dot{a}^2} \right) 
\frac{d \delta}{d a}-\kappa(D) C_1 \frac{1}{a^D \dot{a}^2} \delta
(1+\delta)=\left(\frac{D+1}{D}\right) \frac{1}{1+\delta} \left(\frac{d \delta}{d
a}\right)^2+(1+\delta) (\sigma^2-2\Omega^2) .\label{eqn:deltaD}
\end{equation}
The above equation can be solved in any dimension $D$ if the form of $a(t)$ is given. This gives us a non-singular way to analyse growth of structures in $D$ dimensions, including the case $D=2$. But
in three dimensions we observe that the above equation does not correctly
reduce to equation~(\ref{eqn:3Dddelta}). We may obtain the correct equation in three dimensions by making an additional ansatz, namely, that \begin{equation}
\left(\frac{\dot{a}}{a}\right)^2= \frac{2\kappa(D)}{D} \rho_{\rm bm}.
\end{equation} 
With this, equation~(\ref{eqn:deltaD}) reduces to
\begin{equation}
\frac{d^2 \delta}{da^2}+\left(\frac{6-D}{2a}\right)\frac{d \delta}{d a}-\frac{D}{2
a^2} \delta(1+\delta)=\left(\frac{D+1}{D}\right)\frac{1}{1+\delta} \left(\frac{d \delta}{d
a}\right)^2+(1+\delta) (\sigma^2-2\Omega^2) \label{eqn:ddeltaD}
\end{equation}
Notice that the above equation differs from the earlier equation for $\delta$ in $D$ dimensions, equation~(\ref{eqn:ddelta}), in that there is a factor of $(D-2)$ missing from the coefficient of the $\delta(1+\delta)$ term. Therefore, the above equation does correctly reduce to equation~(\ref{eqn:3Dddelta}) since $(D-2)$ equals unity when $D=3$. When $D=2$, equation~(\ref{eqn:ddeltaD}) gives
\begin{equation}
\frac{d^2 \delta}{da^2}+\frac{2}{a}\frac{d \delta}{d
a}-\frac{1}{a^2}\delta (1+\delta)=\frac{3}{2 (1+\delta)}\left( \frac{d
\delta}{d a}\right)^2+(1+\delta) (\sigma^2-2\Omega^2) \label{eqn:ddelta2}
\end{equation}
On linearising this equation by dropping the $(\sigma^2-2\Omega^2)$
term and the nonlinear terms and solving it we get both a growing mode
as well as a decaying mode for $\delta$.  
The solutions are
\begin{equation}
\delta \propto a^q, \qquad q=(-1\pm\sqrt{5})/2
\end{equation}
(It is interesting to note that one of the power law exponents is the golden ratio).
\par
The Spherical Top Hat (STH) equation in this case turns out to be
\begin{equation}
\ddot{R}=-\frac{GM}{R}+\frac{1}{2} \frac{R}{t^2} \label{eqn:sth2d}
\end{equation}
where $M$ is the constant mass inside a `spherical' shell of radius $R$. 
This equation, unfortunately, has no simple analytic solution.

While this procedure leads to nontrivial results, it has many {\it ad hoc} assumptions and cannot be obtained by taking appropriate limits of Einstein's theory in a systematic manner. Consequently it cannot be applied to numerical
investigations of $2$ dimensional gravity with the confidence that the results will have some implications for the three dimensional case.   

\section{Conclusions} \label{sec:conclusions}
In this paper we have analysed the case of two dimensional gravitational
clustering  starting from a formulation of the $D$ dimensional
Einstein's equations and taking the proper limits. The system of
equations thus arrived at for a $(D+1)$ dimensional universe has been
shown to reduce to the correct equations in three dimensions. But when the
$D\to 2$ limit of these equations is taken, we are forced to  conclude that irrespective of the value of $\kappa (D)$, a consistent two dimensional gravity theory in a cosmological context that supports growth of structures cannot
be constructed.
\par
If $\kappa(2)$ is assumed to be finite,  we observe that  the coefficient in Poisson's
equation goes to zero thus decoupling the potential from the
density.  This implies that perturbations do not
grow but decay in time due to the expansion of the background spacetime.
The alternative which is obtained by using the expression for $\kappa(D)$
given by equation~(\ref{eqn:poisson5}) gives rise to solutions for the
scale factor which are singular and therefore unacceptable.
\par
We have discussed all the ways in which two dimensional gravity may be simulated
including an {\it ad hoc} procedure without a strong foundation which
can give non--singular results as far as structure formation scenarios
in two dimensions are concerned. The results presented in the paper leads us to conclude that the  only way to do a numerical simulation of two dimensional
gravity  is to simulate infinite `needles' in a background
spacetime expanding in three dimensions and consider the `particles'
in the  system to be intersections of the `needles' with any plane
orthogonal to them.
\par

\section{Acknowledgments}
K. Srinivasan and S. Engineer thank CSIR, India for support during the period of work.

\end{document}